\documentstyle[aps]{revtex}
\begin{document}
\title{Towards analytic description of a transition from weak to strong
  coupling regime in correlated electron systems. I. Systematic
  diagrammatic theory with two-particle Green functions}

\author{V.  Jani\v{s}}

\address{Institute of Physics, Academy of Sciences of the Czech Republic,\\
  Na Slovance 2, CZ-18040 Praha 8, Czech Republic}

\date{\today} \maketitle

\begin{abstract}
  We analyze behavior of correlated electrons described by Hubbard-like
  models at intermediate and strong coupling. We show that with increasing
  interaction a pole in a generic two-particle Green function is
  approached. The pole signals metal-insulator transition at half filling
  and gives rise to a new vanishing ``Kondo'' scale causing breakdown of
  weak-coupling perturbation theory.  To describe the critical behavior at
  the metal-insulator transition a novel, self-consistent diagrammatic
  technique with two-particle Green functions is developed. The theory is
  based on the linked-cluster expansion for the thermodynamic potential
  with electron-electron interaction as propagator.  Parquet diagrams with
  a generating functional are derived. Numerical instabilities due to the
  metal-insulator transition are demonstrated on simplifications of the
  parquet algebra with ring and ladder series only. A stable numerical
  solution in the critical region is reached by factorization of singular
  terms via a low-frequency expansion in the vertex function. We stress the
  necessity for dynamical vertex renormalizations, missing in the simple
  approximations, in order to describe the critical, strong-coupling
  behavior correctly.  We propose a simplification of the full parquet
  approximation by keeping only most divergent terms in the asymptotic
  strong-coupling region. A qualitatively new, feasible approximation
  suitable for the description of a transition from weak to strong coupling
  is obtained.
\end{abstract}


\section{Introduction}
\label{sec:introduction}

Tight-binding models of correlated electrons are expected to provide
description of thermodynamic and spectral properties of the underlying
system in weak as well as in strong coupling regimes. We have at our
disposal relatively reliable techniques to describe the two extreme limits
of weak and strong couplings in the archetypal Anderson impurity and
Hubbard lattice models. The weak-coupling regime is governed by a
Hartree-Fock mean field with dynamical fluctuations captured by
Fermi-liquid theory. Extended systems at low temperatures are Pauli
paramagnets with smeared out local magnetic moments. For bipartite lattices
antiferromagnetic long-range order sets on at half filling at arbitrarily
small interaction.  In the strong-coupling regime the Hubbard model at half
filling maps onto a Heisenberg antiferromagnet with pronounced local
magnetic moments and the Curie-Weiss law for the staggered magnetic
susceptibility, at least at the mean-field level. The spectral structure is
dominated by separated lower and upper Hubbard bands and the strongly
correlated system seems to be insulating even in the paramagnetic phase.

In recent years importance of the regime, where the effective Coulomb
repulsion is comparable with the kinetic energy and hence neither very weak
nor very strong, has significantly increased. Unfortunately there is up to
now no adequate method for the description of this transition regime
between the weak and strong-coupling limits. At intermediate coupling
dynamical fluctuations control the low-temperature physics of interacting
electrons and neither weak-coupling nor atomic-like perturbation theories
are adequate.  In this nonperturbative regime we expect the Kondo regime to
set on in the Anderson impurity model \cite{Kasuya85} and eventually
breakdown of the Fermi-liquid regime and metal-insulator transition in the
Hubbard model \cite{Hubbard64}.

New progress in the construction of dynamical mean-field theories via high
spatial lattice dimensions pushed forward our possibilities to investigate
the transition region between weak and strong coupling \cite{Georges96a}.
Although the spatial fluctuations are reduced in high dimensions, there is
no restriction on quantum dynamical fluctuations important at intermediate
coupling. Particularly the quantum Monte-Carlo technique in $d=\infty$
revealed a number of new features of the (disordered) Hubbard model at
finite temperatures \cite{Jarrell92,Rozenberg92,Georges92c,Ulmke95}.
However, the quantum Monte-Carlo is restricted only to relatively high
temperatures. That is why an analytic method based on second-order
perturbation theory, called IPT, was used to study the transition from weak
to strong coupling at zero temperature \cite{Georges92a}. IPT at half
filling reproduces well the finite-temperature Monte-Carlo data and
contains the weak-coupling (up to $U^2$) and atomic solutions as exact
limits.  Contrary to Hubbard-III-like theories it reproduces Fermi liquid
at weak coupling. A metal-insulator transition was found at
$U_c\approx6/\pi\nu$, where $\nu$ is the DOS at the Fermi level
\cite{Zhang93,Georges93b,Rozenberg94}.  Although IPT is an analytic theory,
its solution can be reached only via iterations that depend on the initial
value of the self-energy. When we start with $\Sigma^{(0)}(z)=0$ the
iterations converge for weak interaction to a metallic solution. If we
start with $\Sigma^{(0)}(z)=\Sigma_{at}(z)=U^2/4z$ we end up with an
insulating solution for sufficiently strong interaction. The IPT scheme
fails to converge close to the metal-insulator critical point.  Another
reasoning had to be used to demonstrate the existence and character of the
metal-insulator transition \cite{Moeller95}.

At present IPT seems to be the only analytical-numerical scheme
interpolating between Fermi liquid at weak coupling and paramagnetic
insulator at strong coupling in a frustrated system, where the
antiferromagnetic order is suppressed. However, the solution is numerically
unstable in the most interesting region around the critical interaction.
The quantitative details of the critical behavior at the metal-insulator
transition and the way Fermi-liquid theory breaks down remain unaccessible.
IPT is not a conserving theory and the two-particle functions, important
for the description of inter-particle correlations, are not reliable in the
IPT scheme.  A systematic, consistent, and controllable theory delivering
quantitative description of the critical behavior at the metal-insulator
transition is still missing.

The aim of this paper is to fill up this gap and to develop a systematic
theory interpolating between Fermi-liquid and local-moment regimes suitable
for the description of the critical behavior at the metal-insulator
transition of the Hubbard model or the Kondo behavior in the
single-impurity Anderson model. We start with weak-coupling,
self-consistent perturbation theory and propose a scheme how to extrapolate
it to strong coupling.  Contrary to IPT we concentrate on the two-particle
Green functions showing divergence at the metal-insulator transition. We
develop a systematic renormalized perturbation theory based on parquet-type
diagrams with nontrivial renormalizations of two-particle functions.
Especially we demonstrate the necessity for dynamical vertex
renormalization whenever we approach a critical point. A theory with
sufficient vertex renormalization is derived to describe the critical
region of the metal-semimetal transition with condensation of bound
electron-hole pairs.  It is a simplified version of the parquet algebra
keeping only most divergent contributions to the two-particle scattering
function diverging at the critical point.  It strongly renormalizes
scatterings within singlet electron-hole pairs and leaves inter-pair and
triplet correlations unrenormalized.  To stabilize the numerical solution
in the critical region the two-particle pole is factorized in the spirit of
the renormalization group.  The singular contributions can then be
integrated analytically. The resultant new equations allow for a
numerically stable solution even in the critical region of the condensation
of bound electron-hole pairs.  Such a method of numerical evaluation is
asymptotically exact in the critical region of effective impurity models
within the chosen approximation and makes tractable otherwise inaccessible
diagram sums renormalizing the coupling constant far beyond simple ladder
or ring series.

The layout of the paper is as follows. Section II analyzes the reasons why
numerical solutions of sample models break down when the electron-electron
interaction becomes strong.  A systematic perturbation theory for
intermediate and strong coupling using two-particle functions is expounded
in Sec.~III.  Simple approximations of ring and ladder diagrams are
rederived as examples in Sec.~IV. These simple approximations are used in
Sec.~V to demonstrate the existence of poles in two-particle functions
impeding direct numerical continuation of weak-coupling solutions to the
strong-coupling limit.  Section VI demonstrates the necessity for vertex
renormalizations in self-consistent theories around critical points in
order to comply with a Ward identity binding charge renormalizations
(vertices) to mass renormalizations (self-energy).  A feasible
approximation within parquet algebra with self-consistent vertex
renormalizations suitable for the description of the transition from weak
to strong coupling is derived in Sec.~VII. Last Section VIII brings
conclusions.

\section{Correlated 
  electrons at intermediate coupling}
\label{sec:correlated}

To study correlated electron systems at intermediate coupling
quantitatively we will basically work parallelly with two sample models:
the single-impurity Anderson and the lattice Hubbard models. The former one
will serve as a test case for comparison of developed approximations with
the exact Bethe-ansatz or numerical renormalization-group solution.  The
latter model will explicitly be studied only at the mean-field level, i.e.
in infinite spatial dimensions where spatial fluctuations are suppressed
\cite{Metzner89}. This enables us to cover both the model situations within
one theoretical framework.

We start with the underlying Hamiltonians
\begin{mathletters}\label{eq1}
\begin{eqnarray}
  \label{eq1a}
  \widehat{H}_{SIAM}&=&\sum_{{\bf k},\sigma}\epsilon({\bf k})
  c_{{\bf k}\sigma}^\dagger c_{{\bf k}\sigma}^{\phantom{\dagger}} +
  \varepsilon_f \sum_\sigma f_\sigma^\dagger f_\sigma +Uf_\uparrow^\dagger
  f_\uparrow^{\phantom{\dagger}} f_\downarrow^\dagger
    f_\downarrow^{\phantom{\dagger}} 
  + \sum_{{\bf k},\sigma}\left[V_{{\bf k}}c_{{\bf k}\sigma}^\dagger
  f_\sigma +V_{{\bf k}}^*f_\sigma^\dagger
  c_{{\bf k}\sigma}^{\phantom{\dagger}} \right]
\end{eqnarray}
for the single-impurity Anderson model and with
\begin{equation}
  \label{eq1b}
  \widehat{H}_{HM}=\sum_{{\bf k},\sigma}\epsilon({\bf k})
  c_{{\bf k}\sigma}^\dagger c_{{\bf k}\sigma}^{\phantom{\dagger}} +
  U\sum_{{\bf i}} c_{{\bf i}\uparrow}^\dagger
  c_{{\bf i}\uparrow}^{\phantom{\dagger}} c_{{\bf i}\downarrow}^\dagger
  c_{{\bf i}\downarrow}^{\phantom{\dagger}}
\end{equation}
\end{mathletters}
for the Hubbard model. In the former case we integrate out the conduction
electrons and transform the model onto a single-site theory with dynamical
localized electrons. The interaction dependent partition sum has form of a
local functional integral
\begin{mathletters}\label{eq2}
\begin{equation}
  \label{eq2a}
  {\sf Z}_{Loc}=\int{\cal D}f{\cal D}f^*
  \exp\left\{\sum_{\sigma,n}f_{\sigma n}^*{\cal G}_{\sigma n}^{-1}
    f_{\sigma n} -U\int_0^\beta d\tau f_\uparrow^*(\tau)f_\uparrow(\tau)
    f_\downarrow^*(\tau)f_\downarrow(\tau)\right\} 
\end{equation}
where
\begin{eqnarray}
  \label{eq2b}
  {\cal G}_{\sigma n}^{-1}=i\omega_n+\mu_\sigma-\varepsilon_f
  -V^2G_\sigma(i\omega_n) \; &,\;\;\; &G_\sigma^{(0)}(z)=\int d\epsilon
  \rho(\epsilon)[z+\mu_\sigma-\epsilon]^{-1}  
\end{eqnarray}
is the effective dynamical propagator of the localized electrons and
$\mu_\sigma=\mu+\sigma h$ . The variables $f,f^*$ are Grassmann ones and
$\rho$ is the density of states of the conduction electrons.

The Hubbard model in infinite dimensions, i.e. at the mean-field level can
be mapped onto an impurity problem self-consistently embedded in an
averaged medium \cite{Janis91,Georges92a}. The partition sum can then be
transformed onto a local functional integral identical with (\ref{eq1a})
where only the effective dynamical propagator is defined as
\begin{eqnarray}
  \label{eq2c}
  {\cal G}_{\sigma n}^{-1}=G_{\sigma n}^{-1}+\Sigma_{\sigma n}
  \:&,\;\;\;& G_{\sigma n}=\int d\epsilon\rho(\epsilon)
  [i\omega_n+\mu_\sigma-\Sigma_{\sigma n}-\epsilon]^{-1}
\end{eqnarray}
with $\Sigma_{\sigma n}$ as the self-energy to be determined from an
equation
\begin{eqnarray}
  \label{eq2d}
  G_{\sigma n}=\frac 1{{\sf Z}_{Loc}}\int{\cal D}f{\cal D}f^* f_{\sigma
    n}^*f_{\sigma n}
  \exp\left\{\sum_{\sigma,n}f_{\sigma n}^*{\cal G}_{\sigma n}^{-1}
    f_{\sigma n} -U\int_0^\beta d\tau f_\uparrow^*(\tau)f_\uparrow(\tau)
    f_\downarrow^*(\tau)f_\downarrow(\tau)\right\} \, .
\end{eqnarray}
\end{mathletters}
The physics of both the sample systems is determined completely from the
effective local but dynamical partition sum ${\sf Z}_{Loc}$. Here the
unperturbed propagator is ${\cal G}$ and the only dynamical variable is the
imaginary time $\tau\in(0,\beta)$ or Matsubara frequency $i\omega_n$.

Although the mathematical representation for the impurity Anderson and the
mean-field Hubbard partition sum is formally identical, physical behavior
of these models at intermediate coupling is expected to be different.  The
impurity model remains local Fermi liquid up to infinite interaction
strength \cite{Nozieres74}. At half filling a quasiparticle peak forms
around the Fermi energy ($E_F=0$) in the density of states of the local
dynamical electrons the width of which is determined by a Kondo
(dimensionless) scale
\begin{equation}
\label{eq3} 
 \Delta=\exp\{-\pi U\nu\} 
\end{equation}
and a typical energy of the model (e.g. the bandwidth of the conduction
electrons). It is the existence of this Kondo scale incomparably smaller
than other scales of the underlying Hamiltonian that makes perturbation
expansion fail at intermediate coupling. Except for asymptotically exact
solutions in the low-energy limit such as Bethe-ansatz \cite{Tsvelick83} or
numerical renormalization group \cite{Wilson75}, there is no other reliable
approximate scheme being able to reproduce the Kondo strong-coupling
asymptotics (\ref{eq3}) \cite{note0}.

Recent studies on the Hubbard model in $d=\infty$ show that the Kondo scale
of the impurity model should vanish at a critical interaction ($U_c\approx
6\pi/\nu$) due to the additive self-consistency between the ``unperturbed''
propagator ${\cal G}$ and the self-energy
\cite{Zhang93,Georges93b,Rozenberg94,Moeller95}. The electronic system
undergoes a metal-insulator transition. Fermi-liquid theory and
weak-coupling perturbation expansions break down at this transition.
Breakdown of perturbation theory at the Mott metal-insulator transition is
not caused by the exponentially small Kondo scale but by the divergence of
the effective mass determining criticality of the metal-insulator
transition at zero temperature. Effective mass is connected with the real
part of the self-energy $m^*/m\propto 1-\Sigma^{\prime}$ where
$\Sigma^{\prime}=\partial \Sigma(\omega)/\partial\omega|_{\omega=0}$.
Divergence of the effective mass also causes vanishing of an energy scale.
This energy scale is determined by frequencies $\omega_\pm$ at which the
real part of the self-energy reaches its extremum closest to the Fermi
energy. These frequencies determine an energy interval around the Fermi
level within which Fermi liquid and the quasiparticle picture is
applicable. Fig.~1 shows $\mbox{Re}\Sigma(\omega)$ calculated within
self-consistent ring diagrams. It is clearly seen that $\omega_\pm\to0$
with increasing interaction strength. Vanishing of the difference between
$\omega_+>0$ and $\omega_-<0$ and arising of a jump in
$\mbox{Re}\Sigma(\omega)$ then lead to numerical instabilities and
breakdown of iteration schemes to self-consistent approximations at
intermediate coupling, disregarding whether they can or cannot describe the
metal-insulator transition \cite{Janis96a}.

At intermediate coupling the two investigated models have a vanishing small
scale in common. It causes numerical instabilities in self-consistent
approximations and breakdown of iteration schemes \cite{Janis95}. It
hinders analytic continuation of weak-coupling approximations to the
strong-coupling limit.  To succeed in quantitative description of
intermediate coupling it is necessary to treat the vanishing ``Kondo''
scale separately from the other scales determined by the underlying
Hamiltonians. To do that we must understand the origin of this scale. We
will demonstrate in this paper that the vanishing scale reflects
criticality of a two-particle (actually electron-hole) correlation
function.  It means that a two-particle Green function at intermediate
coupling develops a pole at the Fermi energy. It is known from the early
papers on the Kondo problem that the Kondo scale or the Kondo temperature
is connected with an RPA pole in the magnetic susceptibility
\cite{Yosida65}.  It was pointed out recently \cite{Zhang93} that also at
the metal-insulator transition the local, dynamical susceptibility shows
divergence at the Fermi energy.  In order to get the vanishing scale at
intermediate coupling under control a systematic renormalized perturbation
theory based on approximations for two-particle Green functions is to be
used.

\section{Systematic               
  self-consistent approximations for two-particle Green functions}
\label{sec:systematic}

To develop systematic approximations for one-particle Green functions we
use Dyson's equation enabling us to consider explicitly only one-particle
irreducible diagrams contributing to the self-energy. The situation is more
complicated at the level of two-particle functions. Although we can use a
Bethe-Salpeter equation to extract two-particle irreducible diagrams, we
have three possibilities to do it. We have three channels for multiple
two-particle scatterings defining three types of two-particle
irreducibility. They are the electron-hole (e-h), electron-electron (e-e),
and interaction (U) channels schematically drawn in Fig.~2. Different
channels mean different rearrangements of perturbation expansion. The
chosen irreducible diagrams are summed first and the remaining reducible
ones are summed via a Bethe-Salpeter equation in the end. If perturbation
series converges all rearrangements must lead to the same result.

When studying two-particle functions we need to know the one-particle
self-energy. Hence perturbation theory for two-particle functions cannot be
developed without a parallel expansion for the self-energy.  To approximate
the one and two-particle functions simultaneously we use perturbation
expansion with two-particle functions as a means for developing
controllable comprehensive approximations for a generating thermodynamic
potential. All the physical quantitities will then be determined
consistently from the generating functional via functional derivatives.

As a first step we decide which is the relevant two-particle function for
our purpose to describe intermediate and strong coupling.  It should be a
function approaching a pole with increasing interaction. Such a function
must be related to the interaction (two-particle) part of the underlying
Hamiltonian. The natural generalization of the Hubbard interaction is the
following two-particle function
\begin{equation}
\label{eq5} 
{\cal C}_{{\bf ij}}(\tau )=\langle \widehat{n}_{{\bf i}\uparrow }(\tau ) 
\widehat{n}_{{\bf j}\downarrow }(0)\rangle -\langle \widehat{n}_{{\bf i} 
  \uparrow }(\tau )\rangle \langle \widehat{n}_{{\bf j}\downarrow 
  }(0)\rangle       
\end{equation}
measuring correlations between spin up and spin down densities at different
times and different lattice sites. The brackets denote thermal average.
Fourier transform ${\cal C}({\bf q},i\nu _m)$ of the above function is a
two-particle function approaching a pole when the Hubbard interaction
increases. Note that $\nu _m$ are bosonic Matsubara frequencies. Due to the
space and time translational invariance of the Hamiltonian, $\mbox{Im}{\cal
  C}({\bf q},0)=0$ and the real part has definite sign, namely
\begin{eqnarray}
  \label{eq6}
  {\cal C}({\bf q},0)\propto -UX_\uparrow({\bf q})X_\downarrow({\bf q})<0
  &\;\;\mbox{with}\;\;\; X_\sigma({\bf q})={\cal N}^{-1} 
  \sum_{\bf k} \frac{f(\epsilon({\bf k})-\mu_\sigma) - f(\epsilon({\bf
      k+q})-\mu_\sigma)}{\epsilon({\bf k}) - \epsilon({\bf k+q})} \,.  
\end{eqnarray}
This relation is exactly fulfilled at weak-coupling and can qualitatively
be broken (change of the sign) only if the function ${\cal C}({\bf q},i\nu
_m)$ goes thru a pole.  The pole would indicate breakdown of Fermi liquid.
In the weak-coupling, Fermi-liquid regime we can write
\begin{equation}
\label{eq7} 
 {\cal C}({\bf q},0)=\frac T4\left[ \kappa ({\bf q},0)-\chi ({\bf q},0)\right] <0  
\end{equation}
where $\kappa ({\bf q},i\nu _m)$ is the dynamical compressibility and
$\chi({\bf q},i\nu _m)$ the dynamical magnetic susceptibility.
Eq.~(\ref{eq7}) states that magnetic fluctuations are stronger than charge
fluctuations in the models with the Hubbard interaction. Divergence in the
function ${\cal C}({\bf q},0)$ induces divergence in the magnetic
susceptibility and hence is indication of a magnetic long-range order.

Function ${\cal C}({\bf q},i\nu _m)$ will be approximated by a selection of
classes of Feynman diagrams. Consistency and conservation laws demand that
a two-particle approximation be in concord with an approximation on the
one-particle self-energy. According to Baym and Kadanoff
\cite{Baym61,Baym62} a conserving theory must work with fully renormalized
one-particle propagators. Hence only fully self-consistent approximations
are acceptable at the level of two-particle functions. To construct
consistent approximation schemes we have to build up the generating
functional $\Phi [U,G]$ using the function ${\cal C}({\bf q},i\nu _m)$. To
do that we apply a linked-cluster expansion recently proposed by the author
and J. Schlipf \cite{Janis95}. Its salient feature is that the the
interaction line (non-relativistic photons) is used for edges (diagram
bonds or propagators) and loops of electron Green functions serve as
unperturbed vertices. An analogy between the classical free-energy
functional of the Ising model $W[J,h]$ and the quantum grand potential
$\Omega [U,G^{(0)}]$ is thereby made, i.e. $J\longleftrightarrow U$ and
$h\longleftrightarrow G^{(0)}$.  We use this analogy to construct the
generating functional $\Phi[U,G]$ from the two-particle Green function
${\cal C}({\bf q} ,i\nu _m)$.
 
We represent the grand potential of the Hubbard model in the homogeneous
phase in the following form
\begin{eqnarray}
\label{eq8}
 \frac 1{{\cal N}}\Omega [n;\Sigma ,G]&=&-Un_{\uparrow} n_{\downarrow      
   }-\frac 1{\beta {\cal N}}\sum_{\sigma n,{\bf k}}       
 e^{i\omega_n0^{+}}\left\{ \ln \left[ i\omega _n+\mu _\sigma -\epsilon  
     ({\bf k}) -Un_{-\sigma }-\Sigma_\sigma ({\bf
       k},i\omega_n)\right]\right. \nonumber\\         
& & \left. +G_\sigma ({\bf k},i\omega_n)\Sigma _\sigma ({\bf k},i\omega    
  _n)\right\} + \Phi [U,G]\, ,
\end{eqnarray}
where $n_\sigma,\Sigma_\sigma({\bf k},i\omega_n),G_\sigma({\bf
  k},i\omega_n)$ are variational variables (functions).  We now introduce a
small perturbation $U\rightarrow U+\delta U({\bf q},i\nu_m)$. The function
${\cal C}({\bf q},i\nu _m)$ is defined as a variational derivative of the
generating functional
\begin{equation}
\label{eq9} 
{\cal C}(U;{\bf q},i\nu _m)=\frac{\delta \Phi \left[ U,G\right] }{\delta U({\bf
    q},i\nu _m)}\bigg|_{\delta U=0}\, .       
\end{equation}
Linked-cluster theorem can be used for the inverse transformation
\begin{equation}
\label{eq10} 
 \Phi [U,G]=\frac U{\beta {\cal N}}\sum_{{\bf q}m} \int_0^1d\lambda
 {\cal C}(U\lambda|{\bf q},i\nu _m)\, .      
\end{equation}
Two-particle function ${\cal C}(U\lambda |{\bf q},i\nu _m)$ is understood
as a functional of the full renormalized electron propagator $G({\bf
  k},i\omega_n)$ with the self-energy determined at the interaction
strength $U$.
 
Function ${\cal C}(U|{\bf q},i\nu _m)$ does not directly obey an equation
of motion. It is connected with a general two-particle Green function
${\cal K}_{\uparrow\downarrow}(U|k',k'';q)$ determined from a set of
coupled Bethe-Salpeter equations. To simplify lengthy expressions we
sometimes use a four-vector notation $k=({\bf k},i\omega_n)$ and $q=({\bf
  q},i\nu_m)$ for fermionic and bosonic variables, respectively.  We can
write
\begin{equation}
  \label{eq13}
  {\cal C}(U|{\bf q},i\nu_m)=\frac 1{\beta^2{\cal N}^2}\sum_{k',k''}
  G_\uparrow(k')G_\uparrow(k'+q)G_\downarrow(k'')G_\downarrow(k''+q) {\cal
    K}_{\uparrow\downarrow}(U|k',k'';q)\, . 
\end{equation}
Functions ${\cal K}_{\sigma\sigma'}(k',k'';q)$ can be represented as a sum
of contributions from different two-particle channels
\begin{eqnarray}
  \label{eq11}
  {\cal K}_{\sigma\sigma'}(k,k';q)&=&I_{\sigma\sigma'}
  (k,k';q) + {\cal K}_{\sigma\sigma'}^U(k,k';q) + {\cal
    K}_{\sigma\sigma'}^{eh}(k,k';q) + {\cal
    K}_{\sigma\sigma'}^{ee}(k,k';q)
\end{eqnarray}
where $I_{\sigma\sigma'}$ contains diagrams {\em irreducible} in all the
channels while ${\cal K}_{\sigma\sigma'}^\alpha$ is a sum of diagrams {\em
  reducible} in the particular channel $\alpha$. Each of the reducible
two-particle functions obeys a Bethe-Salpeter equation mixing all the
channels and eventually spins. The equations in the ``horizontal''
electron-hole and electron-electron channels read
\begin{mathletters}\label{eq12}
\begin{eqnarray}
  \label{eq12a}
  {\cal K}_{\sigma\sigma'}^{eh}(k,k';q)=-\frac 1{\beta{\cal
      N}}\sum_{q'}&& \Lambda_{\sigma\sigma'}^{eh}(k,k';q')
  G_\sigma(k+q') G_{\sigma'}(k'+q')\nonumber \\ 
  &&{\cal K}_{\sigma\sigma'}(k+q',k'+q';q-q')\, ,\\ 
\label{eq12b}
   {\cal K}_{\sigma\sigma'}^{ee}(k,k';q)=-\frac 1{\beta{\cal
      N}}\sum_{k''}&&
  \Lambda_{\sigma\sigma'}^{ee}(k,k+k'+q-k'';k''-k)
  G_\sigma(k'') G_{\sigma'}(k+k'+q-k'')\nonumber \\ 
 && {\cal K}_{\sigma\sigma'}(k'',k';k+q-k'')\, .
\end{eqnarray}
The reducible functions in the ``vertical'' or the interaction channel are
determined from equations
\begin{eqnarray}
  \label{eq12c}
   {\cal K}_{\sigma\sigma'}^U(k,k';q)&=&-\frac 1{\beta{\cal
      N}}\sum_{k''\sigma''}\Lambda_{\sigma\sigma''}^U(k,k'';q)
  G_{\sigma''}(k'') G_{\sigma''}(k''+q){\cal K}_{\sigma''\sigma'}(k'',k';q)\, . 
\end{eqnarray}
\end{mathletters}
Here $\Lambda_{\sigma\sigma'}^\alpha(k,k';q)$ are sums of irreducible
diagrams in the $\alpha$-th channel. Equations (\ref{eq12}) have the
structure of parquet diagrams \cite{Weiner70,Jackson82}. To complete the
parquet algebra and to close the equations we must add a definition for the
irreducible functions from each channel.  When no further approximations
are used the parquet equations are completed with a relation between
reducible, ${\cal K}^\alpha$, and irreducible, $\Lambda^\alpha$ functions
\begin{eqnarray}
  \label{eq14}
  \Lambda_{\sigma\sigma'}^\alpha(k',k'';q')&=&{\cal
    K}_{\sigma\sigma'}(k',k'';q)-{\cal
    K}_{\sigma\sigma'}^\alpha(k',k'';q)\, . 
\end{eqnarray}
The generating functional $\Phi [U,G]$ is now fully determined from the
completely irreducible two-particle vertex functions $I_{\sigma\sigma'}
(U\lambda|k',k'';q)$, Bethe-Salpeter equations (\ref{eq11})- (\ref{eq14}),
and the linked-cluster theorem (\ref{eq10}), with (\ref{eq13}). The
self-energy is determined from the saddle-point of the generating
functional $\Omega$ with respect to the fermion propagator $G$.  Note that
the self-energy here measures only dynamical fluctuations beyond the
Hartree approximation.  Even the simplest approximations on the scattering
matrix ${\cal K}_{\uparrow\downarrow}$ lead to nontrivial complex structure
of two-particle functions causing the existence of poles at strong
coupling.

\section{Simple approximations}
\label{sec:simple}

As examples of the general construction we derive a few simple
approximations where we can get rid of the integral over the interaction
strength and calculate all the quantities at the same value of the coupling
constant $U$. In all these approximations the completely irreducible vertex
$I_{\uparrow\downarrow}$ is assumed to be the bare interaction $U$.

\subsection{Ring diagrams}
\label{sec:ring}

We neglect contributions to the two-particle scattering matrix ${\cal
  K}_{\uparrow\downarrow}$ coming from the electron-electron and
electron-hole channels and consider explicitly only the interaction
channel. It means that no multiple scatterings are allowed.  Further on we
neglect the irreducible functions for scatterings of quasiparticles with
the same spin. The two-particle function ${\cal K}_{\uparrow\downarrow}$ in
this approximation then reads
\begin{eqnarray}
  \label{eq15}
  {\cal K}_{\uparrow\downarrow}^{Ring}(U|k',k'';q)&=&\frac
  U{1-U^2X_\uparrow({\bf q},i\nu_m)X_\downarrow({\bf q},i\nu_m)} \,. 
\end{eqnarray}
The integral over the interaction strength can simply be performed leading
to a generating functional
\begin{mathletters}\label{eq16}
\begin{eqnarray}
  \label{eq16a}
  \Phi^{Ring}[U,G]({\bf q},i\nu_m)&=&\frac 1{2\beta{\cal N}}\sum_{{\bf
      q}m}e^{i\nu_m0^+}\ln\left[1-U^2X_\uparrow({\bf 
      q},i\nu_m)X_\downarrow({\bf q},i\nu_m) \right]     
\end{eqnarray}
where
\begin{eqnarray}
  \label{eq16b}
  X_\sigma({\bf q},i\nu_m)&=&\frac 1{\beta{\cal N}}\sum_{{\bf k}n}
  G_\sigma({\bf k},i\omega_n) G_\sigma({\bf k+q},i(\omega_n+\nu_m))  
\end{eqnarray}
is a bubble of two full one-electron propagators. The self-energy can
easily be derived in form of an integral equation
\begin{eqnarray}
  \label{eq16c}
  \Sigma^{Ring}({\bf k},i\omega_n)&=&-\frac{U^2}{2\beta{\cal N}}\sum_{{\bf
      q}m}\left[G_\sigma({\bf k+q},i(\omega_n+\nu_m)) \right.\nonumber \\  
& & \left. +G_\sigma({\bf k-q},i(\omega_n-\nu_m))\right]
   \frac{X_{-\sigma}({\bf q},i\nu_m)}{1-U^2 X_{\uparrow}({\bf q},i\nu_m)
      X_{\downarrow}({\bf q},i\nu_m)}\, .  
\end{eqnarray}
\end{mathletters}
We have a complete set of integral equations the solution of which fully
determines thermodynamic as well as spectral properties of the
self-consistent ring-diagram approximation. If we reduce this approximation
to infinite dimensions, the full extended propagator is replaced by its
local element, we reveal a mean-field version of the ring diagrams recently
studied numerically \cite{Menge91,Janis95}. At the mean-field level the
numerics breaks down at intermediate filling ($U\approx6/\pi\nu$) when the
pole in $C(z)$ is approached \cite{Janis96a}.  We expect the same for the
finite-dimensional version, but the pole will appear at a specific wave
vector ${\bf q}_0$.

\subsection{Ladder diagrams}
\label{sec:ladder}

Next step in developing approximations onto two-particle scattering matrix
${\cal K}_{\uparrow\downarrow}$ is to consider ladders of either multiple
electron-hole or electron-electron scatterings by neglecting the
contributions from the interaction channel. In these ladder approximations
we neglect screening of the interaction due to polarization bubbles of
electron-hole pairs.  Summing the geometric series of the Bethe-Salpeter
equations we obtain for the electron-hole channel
\begin{mathletters}\label{eq17}
\begin{eqnarray}
  \label{eq17a}
  {\cal K}_{\uparrow\downarrow}^{RPA}(U|k,k';q)&=&\frac U{1+\frac
    U{\beta{\cal N}}\sum_{k''}G_\uparrow(k-k'+k'')G_\downarrow(k'')}
\end{eqnarray}
and for the electron-electron channel
\begin{eqnarray}
  \label{eq17b}
  {\cal K}_{\uparrow\downarrow}^{TMA}(U|k,k';q)&=&\frac U{1+\frac
    U{\beta{\cal N}}\sum_{k''}G_\uparrow(k+k'-k'')G_\downarrow(k'')}\, .
\end{eqnarray}
\end{mathletters}
We can again perform the integration in the linked-cluster theorem
(\ref{eq10}) and end up with
\begin{mathletters}\label{eq18}
\begin{eqnarray}
  \label{18a}
  \Phi^{RPA}[U,G]&=&-\frac 1{\beta{\cal N}}\sum_{{\bf
      q}m}e^{i\nu_m0^+}\left\{ UX({\bf q},i\nu_m)-\ln[1+UX({\bf  
      q},i\nu_m)]\right\}   
\end{eqnarray}
for the electron-hole channel and analogously with
\begin{eqnarray}
  \label{18b}
   \Phi^{TMA}[U,G]&=&-\frac 1{\beta{\cal N}}\sum_{{\bf
       q}m}e^{i\nu_m0^+}\left\{ UY({\bf q},i\nu_m)-\ln[1+UY({\bf 
      q},i\nu_m)]\right\} 
\end{eqnarray}
for the electron-electron channel. Here we denoted
\begin{eqnarray}
  \label{18c}
  X({\bf q},i\nu_m)&=&\frac 1{\beta{\cal N}}\sum_{{\bf k}n}G_\uparrow({\bf
    q+k},i\omega_{m+n})G_\downarrow({\bf k},i\omega_n)\, , 
\\ \label{18d}   
  Y({\bf q},i\nu_m)&=&\frac 1{\beta{\cal N}}\sum_{{\bf k}n}G_\uparrow({\bf
    q-k},i\omega_{m-n})G_\downarrow({\bf k},i\omega_n) \, .   
\end{eqnarray}
Note that $\Phi^{RPA}$ generates random-phase approximation with
renormalized propagators, i.e. its renormalized version as introduced by
Suhl in the study of the single-impurity Anderson model \cite{Suhl67}. The
generating functional $ \Phi^{TMA}$ with multiple electron-electron
scatterings describes a renormalized $T$-matrix approximation studied by
Baym and Kadanoff \cite{Baym61}.  The self-energies, respective their
corrections to the Hartree approximation, for ladder diagrams read
\begin{eqnarray}
  \label{18e}
  \Sigma_{\sigma}^{RPA}({\bf k},i\omega_n)&=&-\frac{U^2}{\beta{\cal
      N}}\sum_{{\bf q}m}G_{-\sigma}({\bf k-q},i\omega_{n-m})\frac{X({\bf q},
      i\nu_m)}{1+UX({\bf q},i\nu_m)}\,,  
\\ \label{18f}
  \Sigma_{\sigma}^{TMA}({\bf k},i\omega_n)&=&-\frac{U^2}{\beta{\cal
      N}}\sum_{{\bf q}m}G_{-\sigma}({\bf k+q},i\omega_{n-m})\frac{Y({\bf q},
      i\nu_m)}{1+UY({\bf q},i\nu_m)}\,.     
\end{eqnarray}
\end{mathletters}

The ring and ladder diagrams constitute simplest approximations at the
level of two-particle functions enabling for explicit analytic form of the
generating functional $\Phi$. These approximations are fully
self-consistent and contain geometric series of two-particle bubbles. The
ring diagrams use bubbles of electron-hole pairs of the same spin, while
the RPA and TMA involve bubbles of quasiparticles with opposite spins. Due
to the full self-consistency in these simple approximations it is in
principle possible to find a solution in the whole range of the interaction
strength. It is, however, necessary to stress that these approximations can
be expected reliable only at weak coupling, since only diagrams to second
order are treated exactly here. We use these approximations as samples to
demonstrate breakdown of numerical iteration schemes at intermediate
coupling and the necessity for a separate treatment of the new small scale
connected with zero of the denominator of the two-particle functions from
the self-energy equations.

\section{Two-particle criticality}
\label{sec:two-particle}

To simplify the analysis we leave the general formulation and resort to a
mean-field description. We suppress the momentum convolutions and come from
the full momentum-dependent Green functions to their local parts. Further
on we choose the spin-symmetric solution at zero temperature for the
half-filled band.  These assumptions lead to significant simplifications of
the resultant equations and enable to trace down breakdown of perturbation
theory at strong coupling of the lattice models.

We first analytically continue the equation for the self-energy to real
frequencies and take explicit advantage of the electron-hole symmetry. Then
the generic equation for the self-energy can be written as \cite{Janis96a}
\begin{mathletters}\label{eq19}
\begin{eqnarray}
  \label{eq19a}
  \Sigma(z)=\frac{U^2}2\int_{-\infty}^0\frac{d\omega'}{\pi}&&\left\{\mbox{Im}
    G(\omega'_+)\left[C(\omega'-z)- C(\omega'+z)\right]  
   + \mbox{Im}C(\omega'_+)\left[G(\omega'+z)-G(\omega'-z)\right] 
  \right\}    
\end{eqnarray}
where
\begin{eqnarray}
  \label{eq19b}
  C^{Ring}(\zeta)=\frac{X(\zeta)}{1-U^2X(\zeta)^2}\;,&\;\; &
  C^{RPA}(\zeta)=\frac{X(\zeta)}{1+UX(\zeta)}\;,\;\;
  C^{TMA}(\zeta)=\frac{X(\zeta)}{1-UX(\zeta)}\;.  
\end{eqnarray}
The two-particle bubble $X$ has a representation
\begin{eqnarray}
  \label{eq19d}
  X(\zeta)=-\int_{-\infty}^0\frac{d\omega'}{\pi}\mbox{Im}G(\omega'_+)\left[
    G(\omega'+\zeta)+G(\omega'-\zeta)\right]\, .
\end{eqnarray}
\end{mathletters}
Here $z$ and $\zeta$ stand for complex energies in fermionic and bosonic
functions, respectively.

Troubles with the numerical solution of these equations begin when the
two-particle function $C(\zeta)$ approaches a pole at $\zeta=0$.  In
analogy to the stability condition (\ref{eq6}) this function must fulfill
\begin{eqnarray}
  \label{eq20}
   C(0)&<&0
\end{eqnarray}
in order to keep the effective mass of the interacting electrons positive.
When inequality (\ref{eq20}) is broken, even by intermediate iterations,
the numerical solution fails to converge or a solution on an unphysical
sheet is reached. It is easy to verify that $X(0)<0$ and hence only the RPA
and ring diagrams run into troubles at intermediate coupling.  The TMA is
harmless and can be extrapolated to strong coupling without meeting
numerical problems \cite{Kanamori63}. One can expect this, since due to the
Coulomb repulsion the electron-hole pairs tend to bind. Multiple
scatterings of electrons on electrons do not contribute to this process.
Only the ring and RPA diagrams are relevant for creation of electron-hole
bound states and divergence of the effective electron mass at the 
metal-insulator transition.

The dynamical two-particle function $C$, the local element of the full
correlation function ${\cal C}$, defines at intermediate coupling a new
dimensionless scale
\begin{eqnarray}
  \label{eq21}
  \Delta&=&1+UX(0)\to 0 
\end{eqnarray}
when its pole is approached. To determine the actual strong-coupling
asymptotics of the RPA and ring diagrams it is necessary to calculate the
behavior of the small scale $\Delta$ as a function of the interaction
strength as accurate as possible. On the one hand the existence of this
small scale impedes continuation of weak-coupling solutions to strong
coupling. On the other hand vanishing of this scale at the critical point
of the metal-insulator transition enables us to pick up divergent
contributions in the critical region analytically.  Only the low-frequency
behavior of the two-particle function $C(\omega)$ is relevant in leading
order of diverging effective mass.  We can replace the full frequency
dependence of the denominator of the function $C(\omega)$ by its
low-frequency asymptotics. Such an approach to derivation of the
strong-coupling asymptotics was used by Hamann \cite{Hamann69} in the case
of Suhl's renormalized RPA and recently by the author \cite{Janis96a} for
the ring diagrams. This treatment of the the singular function $C(\omega)$
is asymptotically exact at the critical point with nonintegrable
singularity.

The existing asymptotic expansions for real frequencies can be generalized
to the complex plane in such a way that the analytic structure of the
resultant self-energy is not violated. No spurious poles are generated in
the lower and upper half-planes of complex energies. This property is
essential not to break the sum rules. We hence make the following
low-frequency ansatz for the denominator of $C(\zeta)$ vanishing at the
critical point
\begin{eqnarray}
  \label{eq22}
  1+UX(\zeta)&\doteq&\Delta-i\mbox{sign}(\mbox{Im}\zeta)UX' \zeta\, .  
\end{eqnarray}
We introduced a new parameter $X'$ as the only part of the function
$X(\zeta)$ relevant for the critical behavior of the solution. It is easy
to see that for the simple approximations
\begin{eqnarray}
  \label{eq23}
  X'&=&\pi \nu^2 >0
\end{eqnarray}
where $\nu=\rho(0)$ remains unrenormalized in the Fermi-liquid regime.
Hence the two-particle vertex does not dynamically renormalize in the
low-frequency limit for simple ladder and ring diagrams.

We use ansatz (\ref{eq22}) and simplify the equation for the self-energy by
keeping only leading-order terms due to the singularity of $C(\zeta)$, i.e.
in $|\ln\Delta|$. As a result we obtain two regions for the self-energy.
For finite frequencies, $|z|\gg\Delta/UX'$ the solution reads
\begin{mathletters}\label{eq24}
\begin{eqnarray}
  \label{eq24a}
  \Sigma^{Ring}(z)=\frac 12\Sigma^{RPA}(z)&=&-\frac{G(z)}{2\pi
    X'}\ln\Delta\, .    
\end{eqnarray}
In the limit of small frequencies, $|z|\approx\Delta/UX'$ the self-energy
has another representation
\begin{eqnarray}
  \label{eq24b}
  \Sigma^{Ring}(z)=\frac 12\Sigma^{RPA}(z)&=&-\frac{G(z)}{2\pi X'}\ln\Delta
  +i\widehat{\eta}\frac{\mbox{Im}G(i0^+)}{2\pi X'}\ln[\Delta-
  i\widehat{\eta}UX'z]\, . 
\end{eqnarray}
\end{mathletters} 
Formulas (\ref{eq24}) hold for any external (complex) frequency
$z=\omega+i\eta$ with $\widehat{\eta}=\mbox{sign}\eta$. The solution has
correct Fermi-liquid asymptotics for small as well as for large
frequencies. Moreover, the Herglotz analytic properties of the self-energy
are preserved.

To complete the equation for the self-energy we need to determine the
parameter $\Delta$. We use a representation $X(0)=2\int_{-\infty}^0d\omega'
\rho(\omega') \mbox{Re}G(\omega')$ that can easily be evaluated from the
derived asymptotic form of the self-energy. Only frequencies
$0>\omega>-\sqrt{|\ln\Delta|/\alpha\pi X'}$ are relevant, where
$\alpha=1,2$ for RPA and ring diagrams, respectively. Using (\ref{eq24a})
we finally obtain
\begin{eqnarray}
  \label{eq25}
  X^{Ring}(0)=\sqrt{2}X^{RPA}(0)=\sqrt{\frac{2\pi X'}{|\ln\Delta|}}\, .
\end{eqnarray}
Note that the low-frequency asymptotics of the self-energy, (\ref{eq24b}),
is irrelevant for evaluation of the integral $X(0)$. It, however,
determines the Fermi-liquid, quasiparticle characteristics of the solution.
E.g. the effective mass comes out from (\ref{eq24b}) as
\begin{eqnarray}
  \label{eq26}
  \frac{m^*}m&\propto&\frac{U\nu}{\alpha\Delta}\, .
\end{eqnarray}
We know from earlier studies that Fermi liquid gets less and less
thermodynamically important with increasing interaction, since the width of
the energy interval around the Fermi level, within which Fermi-liquid
(quasiparticle) picture holds shrinks to zero as $\Delta|\ln\Delta|$.
Non-Fermi-liquid contributions, i.e. the frequencies beyond the first
extremum in the real part of the self energy overtake the control over
macroscopic, integral quantities \cite{Janis96a,note1}.

From the definition of the parameter $\Delta$, (\ref{eq21}), and from
(\ref{eq25}) we obtain an asymptotic form for this two-particle scale
\begin{eqnarray}
  \label{eq27}
  \Delta&=&\exp\{-\alpha\pi^2\nu^2U^2\}
\end{eqnarray}
the result derived for the first time by Hamann for the single-impurity
Anderson model \cite{Hamann69}. Our derivation of the strong-coupling
asymptotics holds for the Hubbard model in $d=\infty$ as well. It means
that there is no difference in leading singular order between impurity and
mean-field lattice models at the two-particle critical point.

The exponential asymptotics (\ref{eq27}) lies one order false compared with
the exact Kondo asymptotics (\ref{eq3}). The failure to reproduce the
correct Kondo scale disqualifies the simple ring and ladder diagrams as
appropriate for intermediate and strong couplings. It is evident from the
parallel comparison of two {\em topologically inequivalent} approximations
(rings and ladders) leading to the same low-frequency divergence that in
each approximation important diagrams contributing the same divergence were
neglected. We hence have to go beyond the simple ladder and ring diagrams
in the search for solutions aiming at a reliable description of the
transition from weak to strong coupling.

\section{Necessity 
  for renormalization of the interaction strength}
\label{sec:necessity}

Already early studies following the pioneering papers of Suhl and
co-workers and Hamann, \cite{Weiner70,Beal-Monod70} indicated the necessity
to go beyond the renormalized RPA and to include vertex renormalizations
omitted in the approximation of Suhl. The simple ring and ladder
approximations renormalize only the self-energy and use the unrenormalized
interaction strength $U$ in scattering processes.  It is clear that at
intermediate and strong coupling the actual interaction, the quasiparticles
effectively feel in their mutual scattering events, must be renormalized by
the presence of other quasiparticle pairs. Actually we have to replace the
bare interaction $U$ by a renormalized two-particle function ${\cal
  K}_{\uparrow\downarrow}$ from (\ref{eq12}) projected to the interaction
channel. We will do it systematically in the next section.

However, there is a more basic reason to introduce vertex renormalization
whenever we use mass renormalization. There is a relation between a
two-particle function and a derivative of the one-particle propagator
having the structure of a Ward identity. Baym nad Kadanoff formulated
necessary conditions for a many-body theory to fulfill mass conservation.
A generalized Ward identity connecting an irreducible vertex function with
a variation of the self-energy with respect to an external potential was
thereby derived \cite{Baym62}. Interacting electron system contains not
only mass but also charge. The electron-electron interaction represents
electrostatic energy with $U=e^2/a^*$. In closed systems the electrostatic
potential is generated entirely by the actual charge distribution. The
charge is carried exclusively by particles involved. Hence a redistribution
of mass density must be accompanied by a corresponding charge
redistribution in order not to generate spurious sources of the
electrostatic potential.  This ``charge conservation'', i.e. the entire
electrostatic potential is generated from the charge distribution only, can
be expressed for the Hubbard interaction as a ``Ward identity''
\begin{mathletters}\label{eq28}
\begin{eqnarray}
  \label{eq28a}
  \frac{\partial\Omega(U,\mu_{{\bf i}\sigma})}{\partial U}&=&\sum_{\bf
    i}\left[\frac{\delta^2\Omega}{\delta\mu_{{\bf
          i}\uparrow}\delta\mu_{{\bf i}\downarrow}}+  
    \frac{\delta\Omega}{\delta\mu_{{\bf
          i}\uparrow}}\frac{\delta\Omega}{\delta\mu_{{\bf
          i}\downarrow}}\right]=\sum_{\bf i}\left\{\frac T4[\kappa_{\bf
      ii}-\chi_{\bf ii}]+n_{{\bf i}\uparrow}n_{{\bf  
        i}\downarrow}\right\}    
\end{eqnarray}
with grand potential $\Omega$ defined in (\ref{eq8}) and $\kappa_{\bf ii}$,
$\chi_{\bf ii}$ as the static, local compressibility and susceptibility,
respectively.  We can generalize this static identity using small space and
time inhomogeneous perturbations $U\rightarrow U+\delta U_{\bf
  ij}(\tau,\tau')$ and $\mu_\sigma\to\mu_\sigma+ \delta\mu_{{\bf
    i}\sigma}(\tau)$ to
\begin{eqnarray}
  \label{eq28b}
  \frac{\delta\Phi[U,G]}{\delta U_{\bf ij}(\tau,\tau')}\bigg|_{\delta U=0\atop
    \delta\mu=0}&=& -\frac{\delta
    G_{{\bf ii}\uparrow}(\tau,\tau^+)}{\delta\mu_{{\bf
        j}\downarrow}(\tau')}\bigg|_{\delta U=0\atop\delta\mu=0}\, .      
\end{eqnarray}
\end{mathletters}
Both these identities are fulfilled for an exact solution. The left-hand
side of (\ref{eq28b}) is the fundamental two-particle function for which
the explicit linked-cluster expansion is used. The right-hand side of
(\ref{eq28b}) makes connection to the dynamical susceptibility and
compressibility obtained as a variation of the one-particle Green function.
If (\ref{eq28b}) is violated we cannot refer to inequality (\ref{eq7}).
Identity (\ref{eq28b}) is important for the linked-cluster expansion where
it is used as a fundamental tool for the construction of the generating
functional, i.e. for explicit integration in the linked-cluster theorem
(\ref{eq10}) \cite{Wortis74}. However, even for classical spin systems,
where the situation is much easier, it was impossible to integrate
(\ref{eq28b}) in full and to resolve the generating functional $\Phi$
completely \cite{Wortis69}.

Most of the existing self-consistent approximations renormalize only mass,
i.e. use renormalized one-electron propagators neglecting vertex
renormalization. The Ward identity (\ref{eq28}) is always violated in
self-consistent theories that do not renormalize interaction strength. E.g.
for the Hartree approximation we obtain after a Fourier transform
\begin{eqnarray}
  \label{eq29}
  \frac{\delta G_\uparrow^{Hartree}}{\delta \mu_\downarrow}({\bf
    q},i\nu_m)\bigg|_{\delta U=0\atop 
    \delta\mu=0}&=&UX_\uparrow({\bf q},i\nu_m)X_\downarrow({\bf q},i\nu_m)
  \left[1-U^2X_\uparrow({\bf q},i\nu_m)X_\downarrow({\bf
      q},i\nu_m)\right]^{-1} 
\end{eqnarray}
with the Hartree one-electron propagators. Since $\Phi^{Hartree}\equiv0$,
identity (\ref{eq28b}) is violated by terms proportional to $U$, the small
expansion parameter of the Hartree approximation. Violation of ``charge
conservation'' in the Hartree approximation has no qualitative impact onto
the physics of the solution unless there is a phase transition making the
right-hand side of (\ref{eq29}) diverge. We hence cannot rely upon the
Hartree approximation at critical points. The Hartree phase diagram must be
confirmed by a more advanced approximation complying better with charge
conservation. At least in such a manner that both sides of (\ref{eq28b})
lead to qualitatively the same phase diagram.

To improve the Hartree approximation towards charge conservation we use the
right-hand side of (\ref{eq29}) for the determination of the generating
functional $\Phi[U,G]$ from the linked-cluster theorem. We end up with a
generating functional of the ring diagrams investigated in
Sec.~\ref{sec:ring}. Since neither this approximation contains vertex
renormalizations, we cannot expect Ward identity (\ref{eq28}) to be
fulfilled \cite{Levine68}. The left-hand side of (\ref{eq28b}) for the ring
diagrams is identical with the right-hand side of (\ref{eq29}) where the
Hartree propagators are replaced by those from the ring diagrams. The new
right-hand side of identity (\ref{eq28b}) must be determined from a set of
integral equations and becomes a complicated function of irreducible vertex
functions $\delta\Sigma_\sigma/\delta G_{\sigma'}$. The integral equation
without integration variables can formally be written as
\begin{eqnarray}
  \label{eq30}
 &&\left\{\left[1-G_\uparrow G_\uparrow\frac{\delta\Sigma_\uparrow}{\delta
        G_\uparrow}\right]\left[1-G_\downarrow G_\downarrow
      \frac{\delta\Sigma_\downarrow}{\delta G_\downarrow}\right] -
    G_\uparrow G_\uparrow \left[U+ \frac{\delta\Sigma_\uparrow}{\delta
        G_\downarrow}\right]G_\downarrow G_\downarrow\left[U+
      \frac{\delta\Sigma_\downarrow}{\delta G_\uparrow}\right]\right\}
  \frac{\delta G_\uparrow^{Ring}}{\delta\mu_\downarrow}=\nonumber \\ 
  && -G_\uparrow G_\uparrow\left[U+ \frac{\delta\Sigma_\uparrow}{\delta
        G_\downarrow}\right]G_\downarrow G_\downarrow
\end{eqnarray}

The vertex functions $\delta\Sigma_\sigma/\delta G_{\sigma'}$ contain first
ladder terms with interaction lines renormalized due to electron-hole
polarization bubbles, cf.  Fig.~3. We see that to comply with charge
conservation at least all ladder and ring diagrams must be taken into
account. We can continue iterations beyond the ring diagrams by correcting
the two-particle function ${\cal C}({\bf q},i\nu_m)$ using (\ref{eq30}).
From the new function ${\cal C}({\bf q},i\nu_m)$ we derive the new
self-energy and the vertex functions $\delta\Sigma_\sigma/\delta
G_{\sigma'}$. They again generate new corrections to ${\cal C}({\bf
  q},i\nu_m)$ and the iterations can continue till the convergence is
reached. Already the initial two steps of this iteration scheme indicate
that first the parquet diagrams can be expected to satisfy identity
(\ref{eq28}).

The parquet algebra is very complicated and does not allow for a general
self-consistent (non-perturbative) solution. If all simpler approximations
break identity (\ref{eq28}) we may ask how much the violation of charge
conservation matters.  It is physically {\em inacceptable} if the functions
from the right and left-hand sides of equation (\ref{eq28}) generate {\em
  qualitatively} different phase diagrams and lead to incompatible spectral
properties. We can trust an approximation violating (\ref{eq28}) only if
both sides of (\ref{eq28b}) are qualitatively equivalent or if the next
step in the iteration scheme towards charge conservation corroborates the
results of the chosen approximation. E.g. the phase diagram of the Hartree
approximation can be relied upon if it is confirmed by the ring diagrams.
Phase transitions calculated with the ring diagrams can be seen as
justified if the vertex functions $\delta\Sigma_\sigma/\delta G_{\sigma'}$
do not contain non-integrable singularities that would generate new
singularities in integral equation (\ref{eq30}).

\section{Dipole approximation}
\label{sec:dipole}

It follows from preceding analysis that the full parquet approximation
would be an ideal candidate for a theory interpolating between weak and
strong coupling. It renormalizes the interaction strength to comply with
charge conservation, at least qualitatively. It can be expected to possess
appropriate asymptotic behavior in both the extreme limits. We, however,
also learned that only nonperturbative solutions, treating singularities of
two-particle functions at intermediate coupling analytically, can lead to
numerically stable solutions for arbitrary interaction strength. Parquet
diagrams were studied in connection with the local-moment problem as
improvements of Suhl's approximation. Since no exact solutions to the
parquet equations exist, approximations had to be used. The local ansatz
chosen in \cite{Weiner70} failed to reproduce the Kondo scale at strong
coupling. Neither non-self-consistent analysis of divergent diagrams at
strong-coupling \cite{Beal-Monod70} helped in the selection of the relevant
classes of diagrams. To achieve a feasible extension of the simple
approximations from Sec.~\ref{sec:simple} appropriate for the description
of the metal-insulator transition we must reduce the full parquet algebra.
Since the scatterings in the electron-hole and interaction channels
contribute the same divergence at the critical point we must involve the
two channels self-consistently. To do that we propose a new approximation
scheme backed by physical reasoning selecting relevant scattering processes
leading to the creation of neutral ``dipole'' bound pairs.

In the critical region of the metal-insulator transition we expect that
electrons and holes with opposite spins, i.e. singlets, form pairs of
almost bound states.  It means that if an electron and a hole with opposite
spins get close to each other they tend to stay together for
macroscopically long times.  Probability of mutual scatterings of electrons
and holes with opposite spins is hence much higher than other scattering
events.  The {\em intra-pair} scatterings of the electron and hole from the
binding pairs are pronounced and are mediated by a renormalized interaction
due to the presence of other almost bound electron-hole pairs.  The pairs
have the total charge zero and carry only a dipole moment. The {\em
  inter-pair} interaction is of dipole character and hence weak. There is
no significant renormalization of the electrostatic potential in scarce
interaction processes between pairs or pairs and particles.  The dipole
approximation hence systematically neglects all multiple scatterings where
more than two particles take part.

This intuitive physical picture can mathematically be formulated in terms
of specific approximations in the general parquet algebra (\ref{eq12}).
First of all the three-particle and higher-order irreducible diagrams are
neglected, i.e. the completely irreducible function
$I_{\uparrow\downarrow}=U$. Next, the triplet scatterings are neglected in
favor of singlet scatterings. We can put in the asymptotic limit of the
metal-insulator transition
\begin{mathletters}\label{eq31}
  \begin{eqnarray}
    \label{eq31a}
    \Lambda_{\sigma\sigma}^\alpha(k,k';q)&=&0
\end{eqnarray}
where $\alpha=U, ee, eh$.  This approximation enables us to solve the
parquet equations (\ref{eq12}) in the vertical, interaction channel. The
two-particle function from the vertical channel, i.e. ${\cal
  K}_{\uparrow\downarrow}^v(k,k';q)=U+{\cal
  K}_{\uparrow\downarrow}^U(k,k';q)$ becomes a functional of the
two-particle scattering function from the horizontal channel defined as
\begin{eqnarray}
  \label{eq31b}
    \Lambda_{\uparrow\downarrow}^U(k,k';q)={\cal
      K}_{\uparrow\downarrow}^h(k,k';q)&=& U + {\cal
      K}_{\uparrow\downarrow}^{eh}(k,k';q) + {\cal
      K}_{\uparrow\downarrow}^{ee}(k,k';q)\, .  
\end{eqnarray}
In the critical region of the metal-insulator transition the two-particle
scattering functions ${\cal K}_{\uparrow\downarrow}^{eh}(k,k';q)$ and
${\cal K}_{\uparrow\downarrow}^U(k,k';q)$ show divergence for ${\bf q}={\bf
  q_0}$ and $\nu\to0$.  This divergence is smeared out whenever we build
convolution of these functions connected only by corresponding electron
propagators. Only if we interject bare interaction $U$ between any
convolution of reducible functions from both horizontal and vertical
channels the divergence survives in leading-order also in convolutions.
This restriction reflects the physical fact that multiple scatterings of
bound pairs with other pairs or fermions are negligible at the critical
point where bound electron-hole pairs condense.  Taking this into account
we obtain a solution for ${\cal K}_{\uparrow\downarrow}^v(k,k';q)$ in
closed form and independent of the fermionic four-momenta $k,k'$
\begin{eqnarray}
 \label{eq31c}
 {\cal K}_{\uparrow\downarrow}^v(k,k';q)&=& \Gamma_{\uparrow\downarrow}({\bf
  q},i\nu_m)=\frac U{1-U\Lambda({\bf q},i\nu_m)}\, .
\end{eqnarray}
The pair function $\Lambda({\bf q},i\nu_m)$ is a sum of irreducible
diagrams that cannot be split into two separate components of the same
character by cutting the interaction line. Typical diagrams included and
omitted in the renormalized interaction $\Gamma_{\uparrow\downarrow}$ are
plotted in Fig.~4. The irreducible function $\Lambda({\bf q},i\nu_m)$ is
connected with the two-particle scattering function ${\cal
  K}_{\uparrow\downarrow}^h(k',k'';q)$ analogously to (\ref{eq13})
\begin{eqnarray}
  \label{eq31d}
  \Lambda(q)&=&\frac 1{\beta^2{\cal N}^2}\sum_{k',k''}
  G_\uparrow(k')G_\uparrow(k'+q)G_\downarrow(k'')G_\downarrow(k''+q)
    {\cal K}_{\uparrow\downarrow}^h(k',k'';q)\, . 
\end{eqnarray}
\end{mathletters}

To complete the approximation we must specify the two-particle function
from the horizontal channel. It is determined from equations (\ref{eq12a})
and (\ref{eq12b}) with appropriate irreducible functions.  Since multiple
electron-electron scatterings do not contribute divergent terms in the
critical region of the metal-insulator transition we can neglect these
scatterings completely without affecting leading-order asymptotics. We can
write
\begin{eqnarray}
  \label{eq32}
  \Lambda_{\uparrow\downarrow}^{eh}(k,k';q)=\Gamma_{\uparrow\downarrow}(q)
  &\quad ,\quad & \Lambda_{\uparrow\downarrow}^{ee}(k,k';q)=0\, .
\end{eqnarray}

The parquet algebra is now complete, i.e. (\ref{eq32}) and (\ref{eq12b})
fully characterize ${\cal K}_{\uparrow\downarrow}^{eh}(k',k'';q)$. To
construct the generating functional we rewrite relation (\ref{eq13})
between the two-particle scattering function ${\cal K}$ and the correlation
function ${\cal C}$ in our simplified parquet algebra to
\begin{eqnarray}
  \label{eq33}
  {\cal C}({\bf q},i\nu_m)=\frac{\Lambda({\bf q},i\nu_m)}{1-U\Lambda({\bf
  q},i\nu_m)}\, .
\end{eqnarray}
Using these equations and linked-cluster theorem (\ref{eq10}) we obtain the
functional of the grand potential
\begin{eqnarray}
  \label{eq34}
  \frac 1{{\cal N}}\Omega&& [n;\Sigma ,G]=-Un_{\uparrow} n_{\downarrow      
   }-\frac 1{\beta {\cal N}}\sum_{\sigma n,{\bf k}}       
 e^{i\omega_n0^{+}}\left\{ \ln \left[ i\omega _n+\mu _\sigma -\epsilon  
     ({\bf k}) -Un_{-\sigma }-\Sigma_\sigma ({\bf
       k},i\omega_n)\right]\right. \nonumber\\         
& & \left. +G_\sigma ({\bf k},i\omega_n)\Sigma _\sigma ({\bf k},i\omega    
  _n)\right\} + \frac U{\beta{\cal N}}\sum_{{\bf q}m}e^{i\nu_m0^+}
\int_{0}^1d \lambda \frac{\Lambda[G](U\lambda|{\bf q},i\nu_m)}{1 - U\lambda
  \Lambda[G](U\lambda|{\bf q},i\nu_m)}\, .
\end{eqnarray}
Eqs. (\ref{eq31})-(\ref{eq34}) fully determine the thermodynamic as well as
the spectral properties of the dipole approximation for arbitrary
interaction. They represent an asymptotic solution of the parquet diagrams
in the critical region of the metal-insulator transition, $U\nearrow U_c$,
where electron-hole bound pairs condense.

It is still a tremendous task to solve the above equations in the whole
range of the coupling constant. In the weak-coupling limit it reduces to a
sum of RPA and ring diagrams, i.e. to a FLEX approximation
\cite{Bickers89a}. In the strong-coupling limit the function
$\Gamma_{\uparrow\downarrow}(q)$ develops a pole at the Fermi energy and we
can use the low-energy expansion as demonstrated for simple approximations
in Sec.~\ref{sec:two-particle}. In the critical region of the
metal-insulator transition $U\to U_c$ we can replace in leading order the
function $\Lambda(U\lambda)$ in the grand potential by $\Lambda(U)$. We can
integrate over the interaction strength to obtain the generating functional
in closed form
\begin{eqnarray}
  \label{eq35}
  \Phi[U,G]&\doteq&\frac 1{\beta{\cal N}}\sum_{{\bf q}m}e^{i\nu_m0^+}
  \ln\left[1-U\Lambda[G](U|{\bf q},i\nu_m)\right]\, .
\end{eqnarray}
Note that (\ref{eq35}) differs from the generating functional (\ref{eq16a})
by a factor of two due to the fact that in the critical region we sum up
equivalent contributions from the ring as well as from the ladder diagrams.
The weak coupling limit of (\ref{eq35}) does not hence reproduce
leading-order term in $U$ with the correct weight.

The final step in our approximation in the critical region of the
metal-insulator transition is the low-energy ansatz enabling to treat the
singularity in ${\cal C}({\bf q},\zeta)$ analytically. At the mean-field
level, where no momentum integration appears, we use
\begin{eqnarray}
  \label{eq36}
  \Lambda(\zeta)&\doteq& \Lambda_0 +
  i\mbox{sign}(\mbox{Im}\zeta)\Lambda'\zeta \, .
\end{eqnarray}
We introduced two real and positive ``mean-field'' parameters $\Lambda_0$
and $\Lambda'$ that are to be determined from equations
(\ref{eq31})-(\ref{eq34}) with the above ansatz in the function
$\Gamma_{\uparrow\downarrow}(\zeta)$. Note that there are no closed
equations for $\Lambda_0$ and $\Lambda'$ derivable from a generating
functional. It means that all physical quantities or equations for them
must be determined from the full grand potential (\ref{eq34}),
(\ref{eq35}). The low-frequency ansatz (\ref{eq36}) must be applied only at
the end stage of the evaluation or solution. It is important that contrary
to the simple approximations the dynamical vertex renormalization
$\Lambda'$ is now a nontrivial function of the interaction strength. This
is a minimal condition a theory must fulfill if metal-insulator transition
should be described \cite{Janis96b}.

We are now in a position to perform quantitative analysis of the dipole
approximation in the critical region of the metal-insulator transition. All
the singularities in this approximation are generated by the single
function $\Gamma_{\uparrow\downarrow}(\zeta)$. With the low-frequency
ansatz (\ref{eq36}) we get the singularities under control. It enables us
to follow the solution from the weak-coupling regime up to the critical
interaction where the metal-insulator takes place and to learn the critical
behavior at this transition point in detail. However, a number of technical
steps must be done before we gain the explicit strong-coupling asymptotics
within the dipole approximation.  Quantitative analysis and the results of
the dipole approximation for the single-impurity Anderson model and the
Hubbard mean-field model will hence be presented in the forthcoming paper.

\section{Conclusions}
\label{sec:conclusions}

We analyzed in this paper correlated electron systems described by
single-impurity Anderson and Hubbard models at intermediate and strong
couplings. We concentrated on the half-filled case where the Kodo behavior
in the former and metal-insulator transition in the latter model are
expected. We showed that in both cases a vanishing scale makes numerical
solutions unstable and prevents numerical extrapolation of weak-coupling
approximations to the strong-coupling limit. Hence neither Kondo
strong-coupling asymptotics nor the metal-insulator transition are directly
accessible from the weak-coupling (metallic) side. We demonstrated that the
vanishing, Kondo, scale has its origin in a pole of an appropriate
two-particle Green function. The relevant two-particle function at
intermediate coupling is shown to be a dynamical generalization of the
two-particle part of the underlying Hamiltonian (\ref{eq5}). The pole in
this function signals that singlet bound electron-hole pairs condense and
turn the half-filled metallic solution insulator.

The vanishing Kondo scale at strong coupling makes quantitative description
of this limit difficult. The only way how to get the instabilities
connected with this small scale under control is to treat the Kondo scale
separately from other scales determined by the Hamiltonian or the input
parameters.  To this end we proposed a perturbative scheme enabling direct
access to two-particle functions. We used a quantum version of a
linked-cluster expansion where bonds stand for interaction lines and
vertices for loops of fermionic particles (electrons, holes). With the aid
of the linked-cluster theorem we succeeded in constructing a generating
functional for the parquet diagrams where all three and higher order
irreducible functions are neglected. This approximation is a fully
self-consistent sum of contributions from vertical (ring) and horizontal
(ladder) channels of the full two-particle scattering matrix.

The approximation containing the unrestricted parquet algebra does not
allow for a global nonperturbative solution needed in the strong-coupling
limit.  We analyzed simplified versions of the parquet approximation
summing separately either the ring diagrams or the ladders of electron-hole
and electron-electron multiple scatterings. These approximations, known as
ring approximation, renormalized RPA and TMA, are used to demonstrate the
two-particle instability at intermediate coupling. It was shown that the
ring and electron-hole ladders lead to the same divergence in the effective
quasiparticle mass in the Fermi-liquid regime. We used a low-frequency
ansatz to factorize leading-order singularity of the two-particle
scattering function. It enabled us to treat the vanishing Kondo scale
analytically and separately from the other scales of the interacting system
as demanded from numerical stability. The simple ring and ladder
approximations were evaluated with the low-frequency ansatz in the
strong-coupling limit and the same physically incorrect result for the
rings and electron-hole ladders was obtained. The Kondo scale in both cases
leads to the dependence $T_K\propto\exp\{-\pi^2U^2\nu^2\}$ either in the
single-impurity Anderson model or in the Hubbard model in $d=\infty$.

The failure of the simple ring and ladder approximations to the
two-particle scattering function to reproduce the Kondo scale in the
single-impurity Anderson model or metal-insulator transition in the Hubbard
model lies in the lack of vertex corrections or dynamical renormalization
of the interaction strength. We used a very general argument, interpreted
as Ward identity for charge conservation, connecting variational
derivatives of the generating functional with respect to the spin-dependent
chemical potentials and the interaction strength. This identity, fulfilled
in the exact solution, demands that whenever we renormalize electron mass
(one-electron self-energy) we must adequately renormalize charge
(interaction strength). We showed that first the full parquet diagrams
self-consistently summing ring and ladder diagrams can be expected to
fulfill the Ward identity (\ref{eq28}).

Any reliable approximation suitable for the description of the transition
from weak to strong coupling in interacting electron systems must contain
vertex corrections qualitatively equivalent to those embodied in the
parquet algebra. We analyzed the parquet diagrams in the critical region of
the metal-insulator transition from the metallic side and developed a
dipole approximation. This approximation systematically neglects
nonsingular contributions and keeps only leading-order divergent terms of
the full two-particle scattering matrix from the parquet approximation.
Together with the low-frequency ansatz a feasible approximation was
achieved enabling to study the strong-coupling limit of correlated electron
systems at a qualitatively new level. Moreover, the dipole approximation
obeys Fermi-liquid asymptotics at weak coupling and it can be viewed as a
theory of mean-field character interpolating between the weak and
strong-coupling limits. Since no important, divergent, diagrams from the
parquet diagrams were omitted in the dipole approximation, we may expect
that it will correctly reproduce the Kondo strong-coupling behavior of the
single-impurity Anderson model and eventually the metal insulator
transition in the Hubbard model. Quantitative analysis of the sample models
of interacting electrons with the dipole approximation is left to the
forthcoming paper.

\section*{Acknowledgments}
\label{sec:aknowledgmets}

The work was supported in part by the grant No. 202/95/0008 of the Grant
Agency of the Czech Republic.  I thank Alexander von Humboldt Foundation
for financial support during my stay at the University of Augsburg where
part of this work was performed.

\newpage
\noindent
{\bf Figure Captions} \medskip
\begin{description}
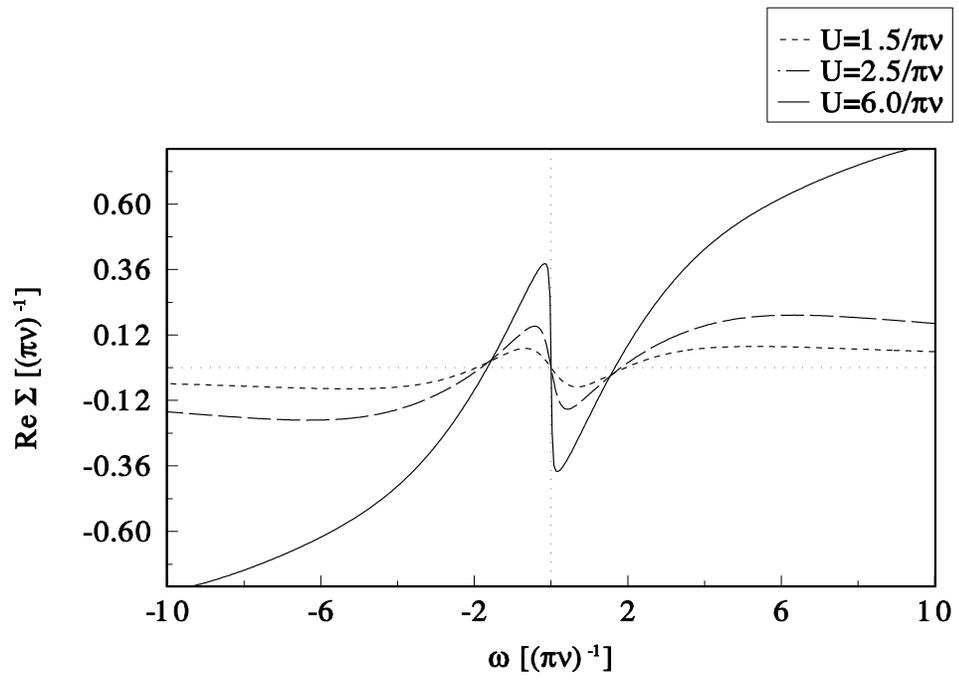
\item[Fig.~1] Real part of the self-energy of the impurity Anderson model
  at half filling and zero temperature calculated for the Lorentzian DOS
  within the ring diagrams from Sec.~\ref{sec:ring}. The first (sharp)
  extrema are clearly seen to approach the Fermi energy, ($\omega=0$), with
  increasing interaction.
  
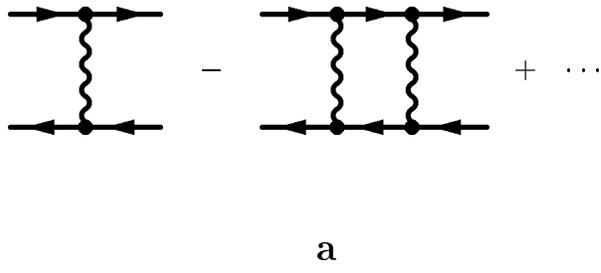
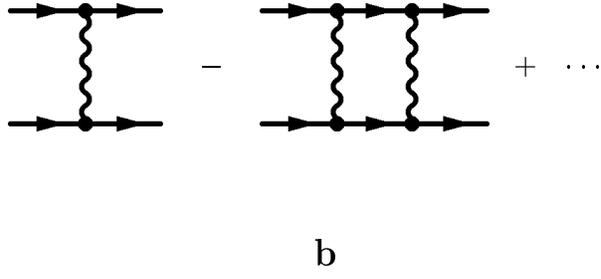
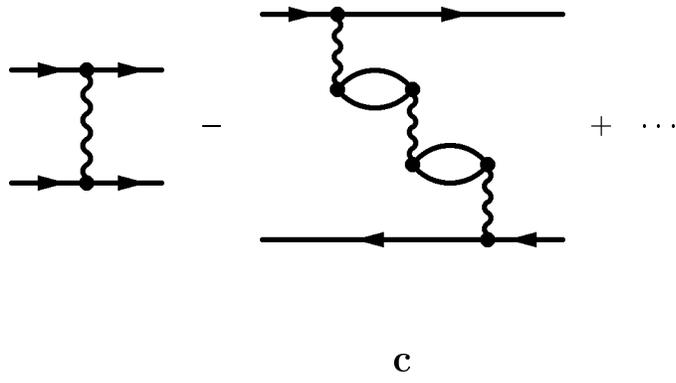
\item[Fig.~2] First few diagrams contributing to the electron-hole, (a),
  electron-electron, (b), and interaction, (c), channels of the
  two-particle scattering function.
  
\item[Fig.~3] New vertex functions generated from functional derivatives of
  the self-energy from the ring diagrams. The double wavy line is the
  interaction between spin up and spin down particles, and the double
  dashed line the same between the particles with the same spin
  renormalized by ring diagrams. These vertex functions are not contained
  in the left-hand side of identity (\ref{eq28b}) of the ring diagrams.
  
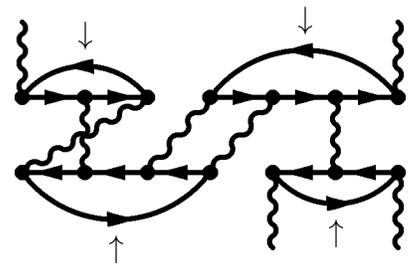
\item[Fig.~4] Typical diagrams contributing to
  $\Gamma_{\uparrow\downarrow}$, (a), (b), where no more than two particles
  take part in multiple scatterings from the irreducible function
  $\Lambda$. Diagram (c) does not contribute to
  $\Gamma_{\uparrow\downarrow}$, since more than two fermion loops are
  multiply connected with the interaction line.
\end{description}


\begin{references}
  
\bibitem{Kasuya85} T.~Kasuya and T.~Saso, editors, \newblock {\em Theory of
    Heavy Fermions and Valence Fluctuations}, SSSSS {\bf 62},
  Springer-Verlag, Berlin, 1985.
  
\bibitem{Hubbard64} J.~Hubbard, \newblock Proc. Roy. Soc. London A {\bf
    281}, 401 (1964) and W.~F. Brinkman and T.~M. Rice, \newblock Phys.
  Rev. B {\bf 2}, 4302 (1970).
  
\bibitem{Georges96a} A.~Georges, G.~Kotliar, W.~Krauth, and M.~Rozenberg,
  \newblock Rev. Mod. Phys. {\bf 68}, 13 (1996).
  
\bibitem{Jarrell92} M.~Jarrell, \newblock Phys. Rev. Lett. {\bf 69}, 168
  (1992).
  
\bibitem{Rozenberg92} M.~J. Rozenberg, X.~Y. Zhang, and G.~Kotliar,
  \newblock Phys. Rev. Lett. {\bf 69}, 1236 (1992).
  
\bibitem{Georges92c} A.~Georges and W.~Krauth, \newblock Phys. Rev. Lett.
  {\bf 69}, 1240 (1992).
  
\bibitem{Ulmke95} M.~Ulmke, V.~Jani{\v{s}}, and D.~Vollhardt, \newblock
  Phys. Rev. B {\bf 51}, 10411 (1995).
  
\bibitem{Georges92a} A.~Georges and G.~Kotliar, \newblock Phys. Rev. B {\bf
    45}, 6479 (1992).
  
\bibitem{Zhang93} X.~Y. Zhang, M.~J. Rozenberg, and G.~Kotliar, \newblock
  Phys. Rev. Lett. {\bf 70}, 1666 (1993).
  
\bibitem{Georges93b} A.~Georges and W.~Krauth, \newblock Phys. Rev. B {\bf
    48}, 7167 (1993).
  
\bibitem{Rozenberg94} M.~J. Rozenberg, G.~Kotliar, and X.~Y. Zhang,
  \newblock Phys. Rev. B {\bf 49}, 10181 (1994).
  
\bibitem{Moeller95} G.~Moeller, Q.~Si, G.~Kotliar, M.~J. Rozenberg, and
  D.~S. Fisher, \newblock Phys. Rev. Lett. {\bf 74}, 2082 (1995).
  
\bibitem{Metzner89} W.~Metzner and D.~Vollhardt, \newblock Phys. Rev. Lett.
  {\bf 62}, 324 (1989).
  
\bibitem{Janis91} V.~Jani\v{s}, \newblock Z. Phys. B {\bf 83}, 227 (1991).
  
\bibitem{Nozieres74} P.~Nozi\`eres, \newblock J. Low Temp. Phys. {\bf 17},
  31 (1974).
  
\bibitem{Tsvelick83} A.~M. Tsvelick and P.~B. Wiegman, \newblock Adv. Phys.
  {\bf 32}, 453 (1983)
  
\bibitem{Wilson75} K.~G. Wilson, \newblock Rev. Mod. Phys. {\bf 47}, 773
  (1975).
  
\bibitem{note0} There are theories based on the non-crossing approximation
  (NCA) that are often used to describe the strong-coupling limit,
  N.~E.~Bickers, \newblock Rev. Mod. Phys. {\bf 59}, 845 (1987). These
  theories, however, do not reproduce the Fermi-liquid regime at zero
  temperature and are hence unsuitable for interpolation between weak and
  strong-coupling limits.
  
\bibitem{Janis96a} V.~Jani\v{s}, \newblock J. Phys. Cond. Matter {\bf 8},
  L173 (1996).
  
\bibitem{Janis95} V.~Jani\v{s} and J.~Schlipf, \newblock Phys. Rev. B {\bf
    52}, 17119 (1995).
  
\bibitem{Yosida65} K.~Yosida and A.~Okiji, \newblock Prog. Theor. Phys.
  {\bf 34}, 505 (1965).
  
\bibitem{Baym61} G.~Baym and L.~P. Kadanoff, \newblock Phys. Rev. {\bf
    124}, 287 (1961).
  
\bibitem{Baym62} G.~Baym, \newblock Phys. Rev. {\bf 127}, 1391 (1962).
  
\bibitem{Weiner70} R.~A. Weiner, \newblock Phys. Rev. Lett. {\bf 24}, 1071
  (1970) and \newblock Phys. Rev. B {\bf 4}, 3165 (1971).
  
\bibitem{Jackson82} A.~D. Jackson, A.~Lande, and R.~A. Smith, \newblock
  Physics Reports {\bf 86}, 55 (1982).
  
\bibitem{Menge91} B.~Menge and E.~M{\"u}ller-Hartmann, \newblock Z. Phys. B
  {\bf 82}, 237 (1991).
  
\bibitem{Suhl67} H.~Suhl, \newblock Phys. Rev. Lett. {\bf 19}, 442 (1967).
  
\bibitem{Kanamori63} J.~Kanamori, \newblock Progr. Theor. Phys. {\bf 30},
  275 (1963).
  
\bibitem{Levine68} M.~Levine and H.~Suhl, \newblock Phys. Rev. {\bf 171},
  567 (1968). Note that here the violation of charge conservation
  (\ref{eq28}) is referred to as non-conservation of spin.
  
\bibitem{Hamann69} D.~R. Hamann, \newblock Phys. Rev. {\bf 186}, 549
  (1969).
  
\bibitem{note1} It was argued that the low-frequency expansion (\ref{eq22})
  may break down in the critical region of the metal insulator transition
  and non-Fermi-liquid contributions have to be taken into account
  \cite{Janis96a}. The non-Fermi-liquid contributions, i.e. from behind the
  first extrema in the real part of the self-energy, are again dominated by
  linear terms leading to the same logarithmic divergences. The only change
  due to non-Fermi-liquid terms is a renormalization of the parameter $X'$.
  
\bibitem{Beal-Monod70} M.~T. B\'eal-Monod and D.~L. Mills, \newblock Phys.
  Rev. Lett. {\bf 24}, 225 (1970).
    
\bibitem{Wortis74} M.~Wortis, \newblock in {\em Phase Transitions and
    Critical Phenomena}, edited by C.~Domb and M.~S. Green, volume~3, p.
  121, eq. 8, Academic Press, London, 1974.
  
\bibitem{Wortis69} M.~Wortis, D.~Jasnow, and M.~A. Moore.  \newblock Phys.
  Rev. {\bf 185}, 805 (1969).
  
\bibitem{Bickers89a} N.~E. Bickers, D.~J. Scalapino, and S.~R. White.
  \newblock Phys. Rev. Lett. {\bf 62}, 961 (1989) and N.~E. Bickers and
  D.~J. Scalapino, \newblock Ann. Phys. {\bf 193}, 206 (1989).
  
\bibitem{Janis96b} V. Jani\v s and R. Tepl\'y, \newblock Czech.\ J. Phys.\ 
  {\bf 46}, 633 (1996)

\end{references}
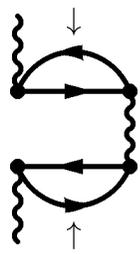
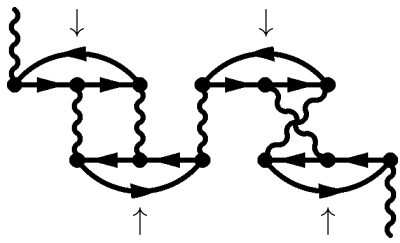
\end{document}